\documentclass[a4paper,11pt]{article}
\usepackage{latexsym,amssymb}
 \usepackage[english]{babel}
 
\usepackage{amssymb,dsfont,amsmath,amsfonts,hyperref,mdframed,mathrsfs,stmaryrd}
\usepackage[active]{srcltx}
\usepackage{slashed}
\usepackage{color}
\usepackage{amsthm}
\usepackage{scrtime}
\usepackage{cite}
\usepackage[normalem]{ulem}
\usepackage{framed}
\usepackage{cancel}
\usepackage{graphicx}
\usepackage{comment}

\makeatletter

\@addtoreset{equation}{section}
\makeatother

\usepackage{amsmath}
\usepackage{amsfonts}
\usepackage{units}
\usepackage{color}
\usepackage{slashed}
\usepackage{parskip}

\def\be{\begin{equation}}
\def\ee{\end{equation}}
\def\ba{\begin{array}}
\def\ea{\end{array}}
\def\bea{\begin{eqnarray}}
\def\eea{\end{eqnarray}}
\def\nn{\nonumber}

\newcommand{\eq}[1]{(\ref{#1})}
\newcommand{\w}[1]{\\[0.#1cm]}

\newcommand{\beqs}{\begin{eqnarray}}
\newcommand{\eeqs}{\end{eqnarray}}

\def\({\left(}
\def\){\right)}

\def\L{ {\cal L}}

\def\mxth{\mathsurround=0pt }
\def\xversim#1#2{\lower2.pt\vbox{\baselineskip0pt \lineskip-.5pt
x  \ialign{$\mxth#1\hfil##\hfil$\crcr#2\crcr\sim\crcr}}}

\renewcommand{\a}{\alpha}
\renewcommand{\b}{\beta}

\newcommand{\e}{\epsilon}

\renewcommand{\L}{\Lambda}
\newcommand{\m}{\mu}
\newcommand{\n}{\nu}

\def\bfone{\relax{\rm 1\kern-.35em 1}}

\newcommand{\bse}{\begin{subequations}}
\newcommand{\ese}{\end{subequations}}

\def\zb{{\bar z}}

\def\vx{{\vec x}}
\def\vE{{\cal E}}

\def\ebp{\overline{\eta_+}\,}
\def\ebm{\overline{\eta_-}\,}

\def\eb{{\bar\epsilon}}
\def\sp{\slashed{\partial}}

\begin{document}

\thispagestyle{empty}

\begin{flushright}\small
MI-HET-837   \\

\end{flushright}

%%%%%%%%%%%%%%%%%%%%%%%%%%%%%%%%%%%%%%%%%

\bigskip
\bigskip

\vskip 10mm

\begin{center}

  {\Large{\bf Singletons in supersymmetric field theories\\[1ex] 
  and in supergravity}}

\vskip 4mm

\end{center}

%%%%%%%%%%%%%%%%%%%%%%%%%%%%%%%%%%%%%%%%

\def\hA{{\widehat A}}
\def\hB{{\widehat B}}
\def\L+{{\widehat\lambda_{+}}}
\def\P-{{\widehat\psi_{-}}}
\def\etam{\overline{\eta_-}}
\def\ep{\overline{\eta_+}}
\def\cm{\widehat{\chi}_{2-}}
\def\hA{\widehat{A}}
\def\hB{\widehat{B}}
\def\ve{{\varepsilon}}

\vskip 6mm

\begin{center}

Henning Samtleben$^{\dagger,\diamondsuit}$ and Ergin Sezgin$^\star$\\
\vskip 8mm

$^\dagger$\, {\it ENSL, CNRS, Laboratoire de physique, F-69342 Lyon, France} \\

\vskip 4mm
 
 $^\diamondsuit$\ {\it Institut Universitaire de France (IUF)}\\
\vskip 4mm

$^\star$\,{\it George P. and Cynthia W. Mitchell Institute \\for Fundamental
Physics and Astronomy \\
Texas A\&M University, College Station, TX 77843-4242, USA}\\
\vskip 4mm

\end{center}

\vskip1.5cm

\begin{center} {\bf Abstract } \end{center}

We consider $N=1$ supergravity coupled to a Wess-Zumino multiplet with a potential such that the resulting AdS energies saturate the unitarity and Breitenlohner-Freedman bounds, respectively. 
Imposing regular boundary conditions that preserve supersymmetry and support finite norms, the spectrum forms  an $OSp(1,4)$ supermultiplet in which the singleton representation is absent. We study  the case in which this boundary condition is relaxed such that the singleton and its superpartners survive and form an indecomposable representation of $OSp(1,4)$. Focusing on the global limit of the model, we carry out its holographic renormalization, in which a singletonic action makes appearance in the boundary action. 

{
\begin{center} 
\em Contribution to Stanley Deser memorial volume ``Gravity, Strings and Beyond"
\end{center}
}

\newpage

\tableofcontents

\section{Introduction}
%%%%%%%%%%%%%%%%%%%%%%%%%%%

The $N=1$ supersingleton multiplet is the shortest unitary representation of  the AdS superalgebra $OSp(1,4)$  consisting of the $SO(3,2)$ irreps %
\be
D(1/2,0) \oplus D(1,1/2)\ ,
\ee
where $D(E_0,s)$ denotes a representation with lowest energy $E_0$ and spin $s$. As is well known, in their field theoretic description these describe boundary states. In view this fact the $N=1$ supersingleton field theory can be formulated as a superconformal field theory of a scalar supermultiplet $(\phi,\chi)$ on the boundary of $AdS_4$. An action for this field theory on the $S^2\times S^1$ boundary of $AdS_4$, and which contains the interactions $\phi^6$ and $\phi^2 {\bar\chi}\chi$, where $\chi$ is a Majorana fermion, can be found in \cite{Nicolai:1988ek}.  The formulation of the supersingleton field theory as a topological field theory of some kind in the bulk, however, is another story. Consider the following action in the (unit radius) AdS bulk 
\be
S_0 = \frac12 \int d^4 x \sqrt{-g} \left( -g^{\mu\nu} \partial_\m \phi\, \partial_\nu \phi +\frac54 \,\phi^2\right) \ .
\ee
The solution of the resulting field equation is described either by the representation $D(5/2,0)$ or $D(1/2,0)$, depending on the boundary conditions imposed. A quantization scheme in which the boundary conditions yield the singleton representation $D(1/2,0)$ was described in detail long ago in \cite{Flato:1980we}.  
It was found that the solution space has an indecomposable structure 
\begin{equation}
    D(1/2,0) \longrightarrow D(5/2,0)
    \;,
    \label{eq:indecomposable}
\end{equation} 
the energy $1/2$ solution ``leaking" into the energy $5/2$ solution under AdS transformations, specifically under dilatations and conformal boosts, cf.~\eq{susyalgebra_bos} below. Singleton modes fall off more slowly at spatial infinity than those described by $D(5/2,0)$. In this sense the singleton modes decouple from the rapidly decreasing “gauge modes” only at the boundary. 

A systematic analysis of the free scalar field  equation in $AdS_4$ was provided in \cite{Starinets:1998dt} for any value of the mass. In the case of $(\Box+\frac54 )\phi=0$, it was highlighted that if one applies the standard definition of the norm, namely $(\phi_1,\phi_2) =i\int d^3x\,\sqrt{-g} g^{0\n} (\phi_1\partial_\n \phi_2-\phi_1\partial_\n \phi_2)$, then the norm of the singleton representation $D(1/2,0)$ diverges, while it is finite for the $D(5/2,0)$ representation. It was also observed in \cite{Starinets:1998dt} that a definition of the norm modified by a multiplication with the factor $(E_0-\frac12)$ gives a finite answer in the case of the singleton. However, the boundary effects encoded in the holographically renormalized action were not taken into account. We will come back to this point in the conclusions, after we describe the holographically renormalization in following sections.

A similar picture arises for spin 1/2 singleton which can be described by the following action in the AdS bulk:
\be
S_{1/2} = \int d^4 x \sqrt{-g} \left({\bar\chi}\gamma^\mu D_\mu \chi -\frac12\,{\bar\chi}\chi \right) .
\ee
In this case, the fermionic singleton representation $D(1,1/2)$ or the regular representation $D(2,1/2)$ arise. 

Turning to supersymmetry, the sum of of the actions $S_0+S_{1/2}$ above does not lend itself to supersymmetrization as can be seen from the failure of the AdS superalgebra on the fermionic field. 
Rather, a minimal $N=1$ supersymmetrization of the model requires the introduction of another scalar with AdS energy $3/2$ which satisfies the Breitenlohner-Freedman (BF) bound \cite{Breitenlohner:1982jf}. 
In analogy with (\ref{eq:indecomposable}) its solution space exhibits an indecomposable structure
\begin{equation}
    D(3/2,0)' \longrightarrow D(3/2,0)
    \;.
    \label{eq:indecomposable32}
\end{equation} 

One may wonder if the singletons may arise in supergravity theories in such a way that while they can be be gauged away in the bulk they would survive as boundary modes. Indeed, the gauged $N=8$ supergravity arising from the $AdS_4\times S^7$ compactification of eleven dimensional supergravity comes with a Kaluza-Klein tower of massive states as well as 8 scalar and 8 spin $1/2$ singletons that form a multiplet of the ${N=8}$ AdS superalgebra $OSp(8|4)$, which, however, can be gauged away (see \cite{Sezgin:2020avr} and references therein.) 
To investigate the prospects of couplings of singletons to supergravity in which they cannot be gauged away, here we consider matter coupled supergravity theories directly in four dimensions which include scalar and spinor fields of singletonic mass. Interestingly, $N=8$ gauged supergravity admits an $N=1$ supersymmetric $AdS_4$ vacuum whose spectrum contains 4 Wess-Zumino multiplets, all of which containing a scalar which carry the representation $D(1/2,0)$ or the regular representation $D(5/2,0)$ depending on the imposed boundary conditions, and a (pseudo)scalar field that carries the regular representation $D(3/2,0)$ \cite{Fischbacher:2009cj,Fischbacher:2010ec}. Motivated by this phenomenon, in this paper we consider the general coupling of a single Wess-Zumino multiplet to $N=1$ supergravity and seek models that can support singletonic states. 
In general, given $N=1$ supergravity coupled to scalar multiplets that admits a supersymmetric $AdS_4$ vacuum solution, the spectrum around such a vacuum contains, in addition to the supergravity multiplet, the following multiplet of states
\be
D(E_0,0) \oplus D(E_0+1/2,1/2) \oplus D(E_0+1,0)\ ,  \qquad E_0>1/2\ ,
\label{WZ}
\ee
for some $E_0$ that depends on the potential, provided that regular boundary conditions that preserve supersymmetry and support finite norms are employed. This is a supermultiplet which forms a unitary irreducible representation (UIR) of the AdS superalgebra $OSp(1,4)$. Note that the singleton representation $D(1/2,0)$ is absent. Our goal, however, is to work with a potential which supports a scalar with singletonic mass, and to explore the consequences of relaxing the standard boundary conditions in such a way that the singleton representation $D(1/2,0)$ arises in the spectrum around the AdS vacuum. 

\begin{figure}[bt]
\center
\includegraphics{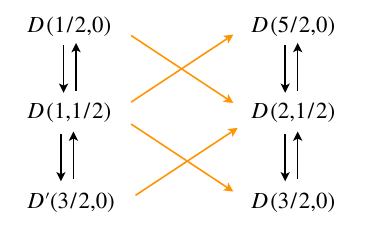}
\caption{\small Indecomposable structure of the supermultiplet. The vertical black arrows show the action of ordinary supersymmetry transformations while the diagonal red arrows show the ``leakage'' under special supersymmetry transformations, cf.~equation (\ref{IDM}) below.}
\label{fig:indecomposable}
\end{figure}

In order to investigate the coupling of singletons to supergravity, we shall consider the $N=1$ AdS supergravity coupled to a single scalar multiplet. In particular, we shall investigate in detail a model in which the scalars of the Wess-Zumino multiplet parametrize the hyperboloid $H^2$, and the potential is chosen such that one of the scalar fields has a singletonic mass. We shall focus on the global limit which captures the special features that arise due to the presence of such mass terms. One of the goals in this paper is to carry out the holographic renormalization procedure by which we study the behaviour of the fields and the action near the boundary, determining the required boundary action to ensure well-definedness of the variational principle and finiteness. Another goal is to investigate the supersymmetry of the system, extending the indecomposable structures (\ref{eq:indecomposable}), (\ref{eq:indecomposable32}), to the full supermultiplet. The resulting structure is summarized in Figure~\ref{fig:indecomposable}, where the red arrows denote the ``leakage'' induced by special supersymmetry transformations.

The key issue of how to establish a finite norm of the singletonic states compatible with physical correlation functions, and the existence of finite conserved charges requires further investigation. We shall comment further on this problem in the concluding section. The resolution of this issue is needed to make progress in discussing the coupling of the singletons to supergravity, and in seeking  a holographic description. 

This paper is organized as follows. In section~2, we recall the form of $N=1$ supergravity coupled to a single Wess-Zumino multiplet with scalars parametrizing a K\"ahler manifold and engineer its  superpotential such that the model admits an AdS vacuum with singletonic scalar mass. In section~3 we consider the global limit of the model around the singletonic vacuum, and determine the boundary terms that arise from the general and supersymmetry variations of the action. In section~4, we determine the boundary expansion of the field equations and their solutions and derive the action of the (conformal) supersymmetry transformations on the boundary data. In section~5, we carry out the holographic renormalization of the model. We determine the boundary terms required for finiteness of the action and analyze supersymmetry of the full renormalized action. We discuss the possible boundary conditions and their relation by Legendre transformation. We comment further on the open problems in section~6.

%%%%%%%%%%%%%%%%%%%%%%%%%%%%%%%%%%%%%%%%%%%%%%%%%%%%%%%%%%%%%%%%%%%%%%%%%%%%%%%%%%
\section{$N=1$ supergravity coupled to Wess-Zumino multiplet}
%%%%%%%%%%%%%%%%%%%%%%%%%%%%%%%%%%%%%%%%%%%%%%%%%%%%%%%%%%%%%%%%%%%%%%%%%%%%%%%%%%

The model is governed by a choice of $K(z,\zb)$ and superpotential ${\cal W}(z)$. While we shall not provide all possible choices for $K$ and ${\cal W}$ that accommodate the coupling of the indecomposable multiplet described above, we shall find a simple example in which this is realized. In this simple example, denoting the complex scalar by $z$, we take the Kahler potential to be
\be
K= -a^2 \log (1-z\zb)\ ,\qquad |z|<1\ ,
\label{K1}
\ee
where we have set $\kappa=1$, and $a$ is an arbitrary real constant, so that the scalar parametrizes the 2-hyperboloid $H^2$ with curvature constant $-4/a^2$.  For an arbitrary superpotential ${\cal W}(z)$, the bosonic part of the action is 
\be
S= \int d^4 x\, \left( \frac12 R - \frac{a^2}{(1-z\zb)^2} \partial_\mu z
\partial^\m \zb -{\cal V}\right) ,
\label{ba}
\ee
where the potential is given by \cite{Cremmer:1982en,Freedman:2012zz} 
\be
{\cal V} =\frac{1}{(1-z\zb)^{a^2}}  \Bigg(  -3|{\cal W}|^2 + \frac{(1-z\zb)^2}{a^2} \Big|\partial_z {\cal W}+\frac{a^2 \zb}{1-z\zb} {\cal W}\Big|^2 \Bigg) \ .
\ee
Thus, looking for supersymmetric AdS vacuum we require that ${\cal V}$ contains an extremum at $z=z_0$, which we can choose to be zero without loss of generality, such that
\be
{\cal W} \ne 0\ , \qquad \Big(\partial_z {\cal W} +\partial_z K\, {\cal W} \Big)_{z=0}\ , \qquad 
\label{susy}
\ee
and that the resulting $2\times 2$ mass matrix for the fluctuations has the mass$^2$ eigenvalues
\be
m^2\,L^2 = \left( -\frac54, -\frac94 \right)\ ,
\label{sing}
\ee
in units of the AdS radius
\be
L = \frac{1}{|{\cal W}|_{z=0}}
\;.
\ee
It follows from  $E_0=\frac32 \pm\sqrt{\frac94+ m^2L^2}$ that $m^2L^2=(-5/4,-9/4)$ gives $E_0=(1/2,3/2)$, with $E_0=1/2$ corresponding to the singleton, and $E_0=3/2$ is massive scalar. Possible choices for $W$ that satisfy these conditions will be systematically analyzed elsewhere. Here we consider the ansatz ${\cal W}=\frac{1}{L}+c_1 z^2$ and taking $c_1$ to be real we find that the conditions \eq{susy}, \eq{sing} yield the solution
\be
{\cal W}= \frac{1}{L}\left(1-\frac14 a^2 z^2\right)\ .
\label{simple}
\ee 
Adding higher order contributions to this superpotential
\be
{\cal W} \longrightarrow {\cal W} +{\cal O}(z^3)\ ,
\ee
will not affect the existence of the singletonic vacuum
but adds higher order interactions to the theory.
The superpotential (\ref{simple}) this gives the potential 
\bea
{\cal V} &=& -\frac{1}{L^2\,(1-z\bar z)^{a^2}}\,
\Bigg[
3 - \frac74 a^2 |z|^2 +\frac{1}{16} a^2(7a^2-8) |z|^4 -\frac{1}{16} a^2(a^2-2)^2 |z|^6 
\nn\\
 && {}\qquad\qquad\qquad\qquad
 - 
 \frac14 a^2 (z - \bar{z})^2 \Big( 1+(2-a^2) |z|^2 \Big)\,\Bigg]\ .
\eea
With this choice, one can check that the conditions \eq{susy} and \eq{sing} are satisfied at the origin of $H^2$, namely at $z=0$, and for any value of $a^2$. One also finds that for $a^2>4$ this potential admits another supersymmetric extremum where the fluctuations form the massive multiplet \eq{WZ} with $E_0$ fixed as follows:  
\be
z_1=\frac{\sqrt{2}}{\sqrt{a^2-2}}<1\ , \qquad m^2\,L_1^2=(4,10)\;\Leftrightarrow\;
E_0=(4,5)
\ .
\ee
The AdS radius in this case is given, in units of the Plank length which has been set to one, by
\be
L_1= \frac{\left(1-|z|^2\right)^{a^2/2}}{|{\cal W}|_{z=z_1}}=2L\left(\frac{a^2-4}{a^2-2}\right)^{(a^2-2)/2}\ .
\ee
Thus, the propagating multiplet has the representations shown in \eq{WZ} with  $E_0=4$. It will be interesting to study solutions of the model that extrapolate between the two extrema one of which harbors the singleton while the other one does not. 

In this paper we shall focus on the holographic renormalization the model in which the complex scalar parametrizes $H^2$ and the superpotential is given by (\ref{simple}). We note, however, that for the complex scalar parametrizing flat space $\mathbb{R}^2$, i.e.\ for the K\"ahler potential
\be
K=a^2\,z\bar{z}\;,
\ee
a suitable superpotential is given by
\be\qquad
{\cal W}=\frac1{L}\left( 1+\frac14\,a^2\,z^2 \right)\ ,
\label{W1a}
\ee
such that the scalar potential takes the form
\be
{\cal V}=\frac{1}{16\,L^2} e^{a^2 z\bar z} 
\left(a^6 \,|z|^6+9\, a^4\, |z|^4+12\, a^2\, |z|^2-48-4\,a^2\,(z-\bar z)^2\,(1-a^2|z|^2)\right)
.
\ee
On the other hand, taking the complex scalar to parametrize $\mathbb{CP}^1$, i.e.\ for the K\"ahler potential
\be
K=a^2\,{\rm ln}(1+z\bar{z})\;,
\ee
the same superpotential (\ref{W1a})also admits a supersymmetric AdS vacuum with a scalar of singletonic mass.
In this case, the scalar potential is given by
\bea
{\cal V} &=& -\frac{(1-z\bar z)^{a^2}}{L^2}\,
\Bigg[
3 - \frac34 a^2 |z|^2 -\frac{3}{16} a^2(3a^2+8) |z|^4 -\frac{1}{16} a^2(a^2+2)^2 |z|^6 
\nn\\
 && {}\qquad\qquad\qquad\qquad
 - 
 \frac14 a^2 (z - \bar{z})^2 \Big( -1+(2+a^2) |z|^2 \Big)\,\Bigg]\ .
\eea

%%%%%%%%%%%%%%%%%%%%%%%%%%%%%%%%
\section{The global limit}
%%%%%%%%%%%%%%%%%%%%%%%%%%%%%%%%

From here on we shall consider the global limit of the model around the singletonic vacuum. This limit captures all the special features that arise due to the singletonic mass.

\subsection{Action and supersymmetry transformations}

The supergravity sector can be included along the lines of \cite{Amsel:2009rr}.
Following \cite{Freedman:2016yue}, we find that the globally supersymmetric limit of $N=1$ supergravity coupled to a single chiral multiplet is given by
\be
S = \int d^4 x\, e \Big[ - g_{z\zb} \partial_\mu z\partial^\m \zb  
-\frac12\Big(  g_{z\zb}\, \bar{\chi_L} \slashed{D} \chi_L  +M \bar{\chi_R}\chi_L + h.c.\Big)-V \Big]\ ,
\label{ba1}
\ee
where the Majorana spinor $\chi$ is written as a sum of left and right handed spinors, namely $\chi=\chi_L+\chi_R$, and  
\bea
V &=& g^{z\zb}\Big|\Big( W' + \frac{1}{L}\partial_z K\Big)\Big|^2 -\frac{3}{L} \Big(W+{\bar W}\Big) -\frac{3}{L^2} K\ .
\nn\w2
M &=&   \Big(\partial_z-\Gamma^z_{zz}\Big)\Big( W'+\frac{1}{L} \partial_z K\Big)\ ,
\nn\w2
D_\m \chi_L &=& \left( \partial_\mu +\frac14 \omega_{\mu ab} \gamma^{ab} + \Gamma^z_{zz} \partial_\mu z \right) \chi_L\ ,
\label{vm}
\eea
where $\Gamma^z_{zz} =g^{z\zb}\partial_z g_{z\zb}$. The superpotential $W$ is related to ${\cal W}$ as 
\be
{\cal W} = \frac{1}{L} +W(z)\ .
\label{W}
\ee
The action is invariant under the supertransformations
\bea
\delta z &=&{\bar\e} \chi_L\ ,
\nn\w2
\delta \chi_L &=& \slashed{\partial} z\, \e_R -g^{z\zb}\left({\bar W}' +\frac{1}{L} \partial_\zb K \right)\e_L\ ,
\label{st1}
\eea
where the supersymmetry parameter obeys the Killing spinor equation 
\be
\nabla_\m \e = -\frac{1}{2L} \gamma_\mu \e\ .
\label{ks}
\ee

The field equations resulting from the action \eq{ba1} are
\bea
&& e^{-1}\partial_\mu \left( e \partial^\mu z\right) +\Gamma^z_{zz} \partial_\mu z \partial^\mu z - g^{z\zb} \partial_\zb V =0\ ,
\nn\w2
&& \slashed{D} \chi_L+  g^{z\zb}{\bar M}\, \chi_R =0\ .
\label{eom1}
\eea
For the model that admits the singletonic vacuum solution we have
\be
W=-\frac{a^2}{4L} z^2\ ,\qquad K=-a^2 \log (1-z\zb)\ , \quad |z|<1\ .
\label{tm}
\ee
Substituting these expressions into \eq{vm},  upon defining $z=A+iB$ and setting $L=1$, gives the action where 
\bea
V&=&\frac{a^2}{4}\Big[
7|z|^2+ 2|z|^4 + |z|^6 +
(z-\zb)^2 (1 + 2|z|^2)  + 12 \log(1 - |z|^2)
\Big]
\ ,
\nonumber\\
M &=&
a^2\left(-\frac12
+\frac{B^2-A^2}{(1-|z|^2)^2}
+\frac{2iAB}{(1-|z|^2)^2}
\right)
\ ,
\eea
and the supersymmetry transformations \eq{st1} take the form
\bea
\delta A &=& \frac12 \eb \chi\ ,
\nn\w2
\delta B &=&- \frac{i}{2} \eb\gamma_5 \chi\ ,
\nn\w2
\delta \chi &=& \Big[ \slashed{\partial} A -\frac12 A(1 - |z|^4) +i\gamma_5 \slashed{\partial} B -\frac{i}{2}\gamma_5  B \left( 1-|z|^2\right)\left(3-|z|^2\right) \Big] \epsilon\ .
\eea
The supersymmetry algebra closes into $AdS_4$ isometries, 
\be
\delta \Phi = K^\mu_{AB} \,\partial_\mu \Phi\ ,\quad A,B=1,\dots,5\ ,
\label{iso}
\ee
where $K^\mu_{AB}$ are associate Killing vectors. In view of the fact that in the global limit described above the action becomes proportional to $a^2$, we will set $a^2=1$ without loss of generality in what follows.

\subsection{Boundary terms from general and supersymmetry variations}
%%%%%%%%%%%%%%%%%%%%%%%%%%%%%%%%%%%%%%%%%%%%%%%%%%%%%%%%%%%%

We begin by ensuring that the Euler-Lagrange equation are satisfied with appropriate boundary conditions. The general variation of the action (\ref{ba1}) takes the form
\be
\delta S= \int d^4x\,\left( \delta z\, \vE_z+ \delta \bar{\chi_L} \vE_{\chi_R} + e\nabla_\mu \Big[ 
-g_{z\zb} \delta z\, \partial^\m \zb  + \frac12 \bar{\chi_L}\gamma^\mu \delta\chi_L \Big] \right) +h.c.\ ,
\label{gv}
\ee
where $\vE$ are the  equations of motion. Next, we consider the supersymmetry variation of the action. One finds that 
\be
\delta_\e S = \int d^4x\, e\nabla_\mu \Big[ -g_{z\zb}\bar{\e_R}\chi_L \partial^\mu \zb  +\frac12 g_{z\zb} \bar{\chi_L}\gamma^\mu \slashed{\partial} z \e_R-\frac12\bar{\chi_L}\gamma^\mu \left({\bar W}'+\frac{1}{L} \partial_\zb K\right)\e_L\Big] +h.c.
\ee
The first term comes from the variation of the bosonic kinetic term, and the second one arises from the variation of the fermionic kinetic term. A little simplification gives
\bea
\delta_\e S &=& -\frac12\int d^4x\, e\nabla_\mu \Big\{ \overline{\e_R}\gamma^\mu\Big[
 g_{z\zb}\,  \slashed{\partial} \zb\, \chi_L -\Big( {\bar W}' +\frac{1}{L} \partial_\zb K\Big)\chi_R  \Big]\Big\} +h.c.
\nn\w2
&=& \int d^4 x \,\partial_\m V^\mu\ ,
\label{S1}
\eea
where\footnote{The overall sign for the last group of terms has been corrected.}
\bea
V^\mu &=& -\frac{e}{2(1-|z|^2)^2} {\bar\e} \,\gamma^\mu \big(\sp(A-i\gamma_5 \big) \chi
\nn\w2
&& +\frac{e}{4(1-|z|^2)} \Big[ {\bar\e}\gamma^\m \chi (A+A^3+AB^2) +{\bar\e} i\gamma_5 \chi (-3B+B^3+BA^2) \Big]\ ,
\label{S2}
\eea
for the model based on \eq{tm}.

\section{Boundary analysis}
%%%%%%%%%%%%%%%%%%%%%%%%%%%%%%%

\subsection{Boundary expansion of the field equations and their solutions}
%%%%%%%%%%%%%%%%%%%%%%%%%%%%%%%%%%%%%%%%%%%%%%%%%%%%%%%%%%%%%%%%%

The field equations are given by
\bea
&& \Box A +\frac{2A \big(\partial_\mu A \partial^\mu A-\partial_\mu B \partial^\mu B\big) +4B\partial_\mu A \partial^\mu B}{1-|z|^2}-\frac1{2 a^2}(1-|z|^2)^2 \partial_A V=0\ ,
\nn\w2
&& \Box B +\frac{2B\big( \partial_\mu B\partial^\mu B-\partial_\mu A\partial^\mu A\big) +4 A\partial_\mu A\partial^\mu B}{1-|z|^2}-\frac1{2a^2}(1-|z|^2)^2 \partial_B V=0\ ,
\nn\w2
&& \left(\slashed{\nabla} -\frac12 \right)\chi +\frac{2}{1-|z|^2} \Big[ (A\sp A+B\sp B) +(A\sp B -B\sp A)i\gamma_5\Big] \,\chi = 
\nn\w2
&& 
\qquad\qquad + \Big( 2A^2-3|z|^2+\frac32 |z|^4-2i\gamma_5 AB \Big)\chi
\eea
where $\Box = e^{-1} \partial_\mu \left(e g^{\mu\nu} \partial_\nu \right)$, $\chi_L=\frac12 (1+\gamma_5)\chi$, and $\slashed{\nabla}$ is the AdS covariant derivative. Note that these results are independent of $a^2$. 

Next, we perform a weak expansion in which we also take into account the asymptotic behaviour of the fields near the boundary. As we shall see below, near the boundary we have $A\sim r^{1/2}, B\sim r^{3/2}$ and $\chi\sim r$. Thus we expand $(A,B, \chi)$ equations of motion up to order $(\frac52,\frac52,3)$, respectively.

\bea
&& \left( \Box +\frac54\right) A 
+2A(1+A^2 ) \,\partial_\mu A \partial^\mu A
=
\frac12A^3 
+\frac12 A^5 + \cdots\ ,
\nn\w2
&& \left(\Box +\frac94\right) B 
- 2 B\partial_\mu A \partial^\mu A +4 A\partial_\mu A\partial^\mu B
=
\frac12A^2 B +\cdots\ ,
\nn\w2
&& \left( \slashed{\nabla} -\frac12\right) \chi + 2\left(1+A^2\right) (A\sp A) \chi 
+2\,(A\sp B -B\sp A)i\gamma_5\,\chi
\nn\w2
&=& -A^2 +\frac32 A^4 \chi -2i\gamma_5 AB \chi +\cdots
\label{seom}
\eea
For the near boundary analysis, we shall work with Poincar\' e coordinates, in which case the metric reads
\be
ds^2 =\frac{1}{r^2} \big( \eta_{ab} dx^a dx^b + dr^2\big)\ .
\label{m1}
\ee
In this coordinate system, the $AdS_4$ Killing vectors take the form
\bea
(\xi^a)^\mu &=& \eta^{a\mu}\;,
\nonumber\\
d^\mu &=&\{t,x,y,r\}\ .
\nonumber\\
(k^a)^\mu\partial_\mu &=& -x^a \,d^\mu\partial_\mu + \tfrac12\,(r^2+x^bx_b)\,\eta^{ac} \partial_c\  
\;.
\label{KV_exp}
\eea

Moreover, the various differential operators are given by
\bea
\slashed{\nabla} &=&-\gamma^3 r \partial_r +r \sp_{(3)} +\frac32 \gamma_3 \ ,
\w2
\sp &=&-\gamma^3 r \partial_r +r \sp_{(3)} \ ,
\label{FL}\w2
\Box &=& -2r \partial_r +r^2 \partial_r^2 + r^2 \Box_{(3)} \ ,
\w2
\Box \left( f(\vx) r^n \right)  &=& n(n-3)\, r^n\, f(\vx) +r^{n+2} \Box_{(3)} f(\vx)\ ,
\w2
\Box \Big( f(\vx) r^n \log r \Big) &=& n(n-3)\, r^n (\log r) f(\vx) + (2n-3) r^n f(\vx) 
\nn\w2
&& +r^{n+2}(\log r) \Box_{(3)} f(\vx)\ ,
\eea
where $\Box_{(3)}= \eta^{ab}\partial_a\partial_b = \partial^a \partial_a$ and $\sp_{(3)}= \gamma^a\partial_a$. The components of the spin connection one-form in $AdS_4$ are given by $\omega^{a3}=e^a$ and $\omega^{ab}=0$. In weak field expansion of the fermions, it is natural to work with their chiral projections defined by $\chi_\pm = \frac12(1\pm \gamma_3) \, \chi$. Then we observe that near the boundary $\chi_+ \sim r$ and $\chi_- \sim r^2$. Thus, the fermionic field equations up to and including terms of order three are given by 
\bea
\Big[r\partial_r (r\partial_r -3) +\frac54\Big] A &=& -2r^2A(1+A^2) \big(\partial^a A \partial_a A+\partial_r A \partial_r A\big) +\frac12 A^3+\frac12 A^5+\cdots\ ,
\nn\w2
\Big[r\partial_r (r\partial_r -3) +\frac94\Big] B &=& 2r^2 B\big( \partial^a A \partial_a A+\partial_r A \partial_r A\big)  -4 r^2 A \big( \partial^a A \partial_a B +\partial_r A \partial_r B\big) +\frac12 A^2 B+\cdots\ ,
\nn\w2
( r \partial_r -1)\chi_+   &=& r \sp_{(3)} \chi_- - 2(1+A^2) A (r\partial_r A) \chi_+   
+ A^2\chi_+  -\frac32 A^4 \chi_+ + \cdots\ ,
\w2
(r \partial_r -2)\chi_-  &=& -r \sp_{(3)} \chi_+    -2A(r\partial_r A) \chi_- - 2r  (A\sp_{(3)} A) \chi_+ -A^2\chi_- 
\nn\w2
&& 
 +2\Big[ -A (r\partial_r  B) +B (r\partial_r  A)  -AB \Big] i\gamma_5\chi_+  +\cdots\ ,
\nn
\eea
The near boundary expansion of the solutions of the equations \eq{seom} then takes the form
\bea
A(r,\vx) &=& r^{1/2} A_1  +r^{5/2} \left( A_2 - \frac12 \,
{\rm log}\, r \, \Box_{(3)} A_1 \right)\ ,
\nn\w2
B(r,\vx) &=& r^{3/2} \Big( B_1 + {\rm log}\, r \, B_2\Big) 
-2\,r^{5/2} \,A_1^2 \Big(  B_1-B_2 
+ {\rm log}\, r \, B_2 \Big)\ 
+\frac12 \,r^{5/2}\,A_1\,\overline{\chi_{1+}}\gamma^5 \chi_{1+}
\ ,
\nn\w2
\chi(r,\vx) &=& \chi_{+}(r,\vx) + \chi_{-}(r,\vx)\ ,
\eea
where
\bea
\chi_{+}(r,\vx) &=& r \chi_{1+}
 -\frac54 r^{3} A_1^4\,\chi_{1+}
 + \frac12\,r^{3} \slashed{\partial}_{(3)} \chi_{2-}     +\frac14 r^3\big( 1-2\log\,r\big)\Box_{(3)}\chi_{1+} \ ,
\label{FG}\w2
\chi_{-}(r,\vx) &=&  r^2 \Big( \chi_{2-} 
- {\rm log}\, r \, \slashed{\partial}_{(3)} \chi_{1+}\Big)
-2 r^{3}\,A_1^2\,\chi_{2-} -2r^{3} A_1 \left(\slashed{\partial}_{(3)}A_1\right)  \chi_{1+}
\nn\w2
&& 
-2 r^3A_1^2\Big( 1-{\rm log}\, r \Big) \slashed{\partial}_{(3)} \chi_{1+}
-4i\gamma_5 \,r^{3} A_1\,\Big(  B_1-\frac12\,B_2+ {\rm log}\, r \, B_2
\Big)\,\chi_{1+}\ .
\nn
\eea

\subsection{The supersymmetry transformation rules}
%%%%%%%%%%%%%%%%%%%%%%%%%%%%%%%%%%%%%%%%%%%%%%%%%%%%%%

To study the supersymmetry transformations of the solutions presented above, we need the Killing spinors. The  Killing spinor equation \eq{ks}, it is solved by
\be
\e = r^{-1/2} \eta_- + r^{1/2} \eta_+\ ,\qquad \gamma_3 \eta_\pm = \pm \eta_\pm\ ,
\ee
with coefficients $\eta_\pm (\vx)$ obeying 
\be
\partial_a \eta_- = -\gamma_a \eta_+
\ee
which in turn which implies that $\slashed{\partial}\eta_+=0$ and $\slashed{\partial}\eta_-=-3\eta_+$.
Substituting these expansions into \eq{st1} we find  
\bea
\delta A_1 &=& \frac12 \overline{\eta_-}\, \chi_{1+}\ ,
\nn\w2
\delta\chi_{1+} &=& \slashed{\partial} A_1 \eta_- +i\gamma_5 B_2 \eta_- -A_1 \eta_+\ ,
\nn\w2
\delta B_2 &=&  \frac{i}{2} \ebm \gamma_5 \slashed{\partial} \chi_{1+}\ ,
\nn\w4
\delta B_1 &=& -\frac{i}{2} \ebp \gamma_5\chi_{1+} -\frac{i}{2} \ebm \gamma_5 \chi_{2-}  \ ,
\nn\w2
\delta\chi_{2-} &=& \Big[ i\gamma_5 \slashed{\partial} B_1  -\frac12 \left(\Box A_1 -4A_2 - A_1^5\right)\Big] \eta_-
+\Big[ \slashed{\partial} A_1 - i\gamma_5 \left( B_2+3B_1\right)\Big] \eta_+\ ,
\nn\w2
\delta A_2 &=& \frac12 \overline{\eta_+}\, \chi_{2-}+\frac14 \overline{\eta_-}\,\slashed{\partial}\chi_{2-}+\frac18 \overline{\eta_-} \Big(\Box_{(3)}-5A_1^4 \Big) \chi_{1+}\ .
\label{IDM}\eea
We note that this result can be simplified by defining
\begin{center}
\be
\boxed{
\begin{aligned}
{\mathfrak A}_2=A_2-\frac14\Box_{(3)}A_1+\frac14 A_1^5
\label{A2}
\end{aligned}
}
\ee
\end{center}

upon which $\delta\chi_{2-}$ and $\delta A_2$ are to be replaced by
\bea
\delta\chi_{2-} &=& \Big[ i\gamma_5 \slashed{\partial} B_1  +2\,{\mathfrak A}_2 \Big] \eta_-
+\Big[ \slashed{\partial} A_1 - i\gamma_5 \left( B_2+3B_1\right)\Big] \eta_+\ ,
\nn\w2
\delta {\mathfrak A}_2 &=& \frac12 \overline{\eta_+}\, \chi_{2-}+\frac14 \overline{\eta_-}\,\slashed{\partial}\chi_{2-}
-\frac14\, \overline{\eta_+}\, \slashed{\partial} \chi_{1+}
\ .
\label{IDMB}
\eea

The variation of the components under the AdS isometries (\ref{iso})
is straightforwardly obtained with (\ref{KV_exp}) as
\bea
\delta A_1 &=& \xi^a \partial_a A_1   + \tfrac12\, \lambda_D A_1\ ,
 \quad \delta {\mathfrak A}_2 = \xi^a \partial_a {\mathfrak A}_2   + \tfrac52\, \lambda_D {\mathfrak A}_2
 -\tfrac12\,\lambda_D \Box_{(3)}A_1
 + k^a \partial_a A_1 \ ,
\nonumber\\{}
\delta B_2 &=& \xi^a \partial_a B_2   + \tfrac32\, \lambda_D B_2
 \ ,
\quad
\delta B_1 = \xi^a \partial_a B_1   + \tfrac32\, \lambda_D B_1+  \lambda_D B_2\ ,
 \label{susyalgebra_bos}
\eea
with the parameters satisfying the relations 
\be
\partial_a\xi^b = -\delta_a^b\,\lambda_D\ ,\quad
\partial_a \lambda_D =-2 k_a\ ,\quad
\partial_a k^b = 0\ .
\ee
This shows the indecomposable structure of the transformations already on the the level of the bosonic conformal transformations. 
On the bosonic fields, the supersymmetry transformations (\ref{IDMB}) close into AdS isometries with parameters
\bea
\xi^a = \overline{\eta_{-2}}\,\gamma^a \eta_{-1}\ ,\quad 
k^a = \overline{\eta_{+2}}\,\gamma^a \eta_{+1}\ ,\quad \lambda_D =  2\,\overline{\eta_{+[2}}\, \eta_{-1]} 
\eea
Similarly, on the fermion fields, the supersymmetry transformations
close into AdS isometries acting as
\bea
 \delta \chi_{1+} &=& \xi^a\partial_a \chi_{1+}   +  \lambda_D \chi_{1+} 
 \;,\nonumber\\{}
 \delta \chi_{2-} &=& \xi^a \partial_a \chi_{2-}   + 2\, \lambda_D \chi_{2-} -\tfrac12\,\lambda_D \slashed{\partial}\chi_{1+} 
 + \tfrac32\, \slashed{k}\,\chi_{1+}
 \;,
 \eea
together with Lorentz transformations with parameter
\begin{equation}
\Lambda_{ab}= 2\overline{\eta_{+[1}}\,\gamma_{ab} \eta_{-2]}
\ .
\end{equation}

In obtaining the above results, we have used the Fierz identities
\bea
\eta_{-[1} \overline{\eta_{2]-}} &=& -\tfrac14 \xi^a \gamma_a (1+\gamma_3)\ ,\qquad \eta_{-[1} \overline{\eta_{2]+}} =\tfrac18 \big(  \lambda_D - \tfrac12 \Lambda_{ab}  \gamma^{ab} \big) (1-\gamma_3)\ ,
\nn\w2
\eta_{+[1} \overline{\eta_{2]-}} &=& \tfrac18 \big( \lambda_D - \tfrac12 \Lambda_{ab}  \gamma^{ab} \big) (1+\gamma_3)\ ,
\eea
where $\Lambda_{ab}$ is the Lorentz parameter. We observe that the set of fields $(A_1, B_2, \chi_{1+})$ transform strictly into each other thus forming  the following multiplet:
\bea
\delta A_1 &=& \frac12 \overline{\eta_-}\, \chi_{1+}\ ,
\nn\w2
\delta B_2 &=&  \frac{i}{2} \ebm \gamma_5 \slashed{\partial} \chi_{1+}\ ,
\nn\w2
\delta\chi_{1+} &=& \slashed{\partial} A_1 \eta_- +i\gamma_5 B_2 \eta_- -A_1 \eta_+\ .
\label{m2}
\eea

As such, these fields can be treated as sources and consistent with supersymmetry they can be set to zero:
\be
A_1=0\ ,\qquad B_2=0\ ,\qquad \chi_{1+}=0\ .
\label{bc1}
\ee
The remaining fields, namely $(B_1, A_2, \chi_{2-})$ do not only transform into each other but to also the fields $(A_1, B_2, \chi_{1+})$. Therefore, the full set of fields in \eq{IDM} form an indecomposable supermultiplet as depicted in Figure~\ref{fig:indecomposable} above. In contrast, the analog of \eq{IDM} for fields that do not contain logarithmic terms give decomposable representations; see, for example, \cite{Freedman:2016yue} for the case of scalar fields that have conformal dimensions $\Delta=1$ and $\Delta=2$.

Turning to the supertransformations \eq{m2}, they form the superalgebra $OSp(1,4)$ with (anti)commutator rules 
\begin{align}
\{Q_\a,S_\b\} &=(\gamma^i)_{\a\b} J_i + \e_{\a\b}D \ , 
\label{wz1}\\
\{S_\a,S_\b\} &= -2(\gamma^i)_{\a\b} K_i \ ,
\label{wz2}\\
[K_i,Q_\a] &= \tfrac12 (\gamma_i S)_\a\ ,
\label{wz3}\\
[P_i,K_j] &= -\epsilon_{ijk} J^k +\eta_{ij} D\ .
\label{wz4}
\end{align}

\section{The renormalized action and its Legendre transform}
%%%%%%%%%%%%%%%%%%%%%%%%%%%%%%%%%%%%%%%%%%%%%%%%%%%%%%%%%%%%%%%%

\subsection{Boundary terms required by finiteness of the action}
%%%%%%%%%%%%%%%%%%%%%%%%%%%%%%%%%%%%%%%%%%%%%%%%%%%%%%%%%%%%%%%%%%%%%%%%%%%%%%%%%%%%%

The near boundary expansion allows to determine the divergent part of the action. Defining the regularized action as
\be
S_{\rm reg} =
\int_\varepsilon^\infty \! dr \int dx^3\,{\cal L}
\label{b1} 
\ee
we find
\begin{center}
\be
\boxed{
\begin{aligned}
S_{\rm reg}&=
-\frac12 \int dx^3\left(
\frac{1}{\varepsilon^2}\,A_1^2 +
\frac{1}{\varepsilon}\,A_1^4
+3\,({\rm log}\,\varepsilon)^2\,B_2^2
\right)
%\nonumber
\\
&{}
- {\rm log}\,\varepsilon\, \int dx^3
\left(
3\,B_1B_2+B_2^2+\frac32\,\partial_a A_1 \partial^a A_1 
\right)
+ {\cal O}(\varepsilon^0)
\label{sreg}
\end{aligned}}
\ee
\end{center}
Nicely, all divergent terms can be removed by adding a (divergent) boundary action of covariant counterterms 
\begin{center}
\be
\boxed{
\begin{aligned}
S_{\rm ct,1}&=
\frac12\,
\int dx^3\,\sqrt{-\gamma}\,\left(
A^2+A^4+3\,B^2+\frac{2\,B^2}{{\rm log}\,\varepsilon}
+2\,{\rm log}\,\varepsilon\,\gamma^{mn}\,\partial_m A \partial_n A
\right)
\label{eq:Sct1cov}
\end{aligned}
}
\ee
\end{center}
with the induced metric $\gamma_{mn} =g_{mn}= \varepsilon^{-2} \eta_{mn}$ on the boundary, where $m=0,1,2$, is the curved index. 
As established in the holographic renormalization program~\cite{Bianchi:2001kw}, it is important that
the subtractions of the divergent terms are expressed covariantly 
in terms of the fields $A$, $B$,
living on the regulating hypersurface, rather than their components $A_1, A_2, B_1, B_2$.
As a consequence these counterterms induce finite contributions, explicitly
displayed in the second line of the expansion
\begin{align}
S_{\rm ct,1}&=
\int dx^3\,\Big(
\frac1{2\ve^2}\,A_1^2
+\frac{1}{2\ve}\,A_1^4
+\frac32\,({\rm log}\,\varepsilon)^2\,B_2^2
+{\rm log\,\ve}\,
\Big\{
3B_1B_2+B_2^2
-\frac32\,\partial_a A_1 \partial^a A_1
\Big\}
\nonumber\\
&{}\qquad\qquad
+A_1A_2+\frac32\,B_1^2+2\,B_1B_2
\Big)
.
\label{eq:Sct1comp}
\end{align}

\subsection{Boundary terms from supersymmetry and anomalies}
%%%%%%%%%%%%%%%%%%%%%%%%%%%%%%%%%%%%%%%%%%%%%%%%%%%%%%%%%%%%%%%%%%%%%%%%%%%%%%%%%%%%%

Having obtained a finite action, we can now move on to analyze supersymmetry. As a check of consistency, we can verify explicitly, that all divergent terms in the supersymmetry variation of $S_{\rm reg} + S_{\rm ct,1}$ cancel.
In turn, we
find a remaining 
finite boundary contribution for the supersymmetry variation
\bea
\delta_\e \left( S_{\rm reg} + S_{\rm ct,1} \right) &=&
-\delta_\e 
\int d^3 x\, \sqrt{-\gamma}\, 
\Big\{ 
\frac13\, A_1^6
+\frac12\,
\overline{\chi_{1+}}\chi_{2-} 
\Big\}
\nonumber\\
&&{}
+\int d^3 x\, \overline{\eta_+}\left(
A_1\,\slashed{\partial} \chi_{1+}-i B_2\,\gamma^5\,\chi_{1+} \right)
\;.
\label{ct2}
\eea

The terms in the first line can be removed by adding the covariant finite counterterms

\begin{center}

\be
\boxed{
\begin{aligned}
S_{{\rm ct},2} &= \int d^3 x\, \sqrt{-\gamma}\, 
\Big\{ 
\frac13\, A^6
+\frac14\,\big(
\overline{\chi}\chi+2\,({\rm log} \,\varepsilon)\,
\overline{\chi}\gamma^m\partial_m \chi
\big)
\Big\}
\nonumber\\
&=\int d^3 x\, \sqrt{-\gamma}\, 
\Big\{ 
\frac13\, A_1^6
+\frac12\,
\overline{\chi_{1+}}\chi_{2-} 
\Big\}
\label{b3}
\end{aligned}}
\;.
\ee
\end{center}

\bigskip

which are sufficient to guarantee invariance of the resulting action under ordinary supersymmetry $\eta_-$.
On the other hand, we find that the cancellation of the terms in the second line of \eq{ct2} would require additional finite counterterms of the form
\be
\begin{aligned}
S' 
&=
\int d^3 x\, 
\Big\{ 
-\frac12\,\overline{\chi_{2-}}\chi_{1+}
+2A_1{\mathfrak A}_2-B_1 B_2
\Big\}
\\
&=
\int d^3 x\,  
\Big\{ 
-\frac14\,\sqrt{-\gamma}\,\Big(
\overline{\chi}\chi+2\,({\rm log} \,\varepsilon)\,
\overline{\chi}\gamma^m\partial_m \chi
\Big)
+2A_1{\mathfrak A}_2-B_1 B_2
\Big\}
\;,
\end{aligned}
\label{Sf+}
\ee
of which however the last two terms cannot be written in covariant form. However, we note that $\delta_{\eta_{-}} S'=0$. 

Finally we can add the following two parameter boundary action which is finite and supersymmetric under both $\eta_\pm$ supersymmetries,
\begin{center}
\be
\boxed{
\begin{aligned}
S_{\it singleton} &= \int d^3 x\, \sqrt{-\gamma}\, 
\Big\{ 
\kappa_1 \Big[ \gamma^{mn} \partial_m A\partial_n A +\frac{B^2}{(\log\ve)^2 } + 
\overline{\chi}\gamma^m \partial_m \chi \Big]
\w2
& \qquad\qquad\qquad   + \kappa_2 \Big[ \frac{BA^3}{\log\ve}  + \frac{3i}{4}\,\overline{\chi} \gamma_5 \chi A^2\Big]   \Big\}\ 
\\
&=
\int d^3 x\,  
\Big\{ \kappa_1 \Big[
 \partial_a A_1\partial^a A_1 - B_2^2 + \frac12 \overline{\chi_{1+}}\slashed{\partial} \chi_{1+} \Big]
\w2
& \qquad\qquad\qquad +\kappa_2\Big[ B_2A_1^3+ \frac{3i}{4}\,\overline{\chi_{1+}} \gamma_5 \chi_{1+} A_1^2 \Big] \Big\}\ 
\label{SL}
\end{aligned}}
\ee
\end{center}
where $\kappa_1$   and $\kappa_2$ are arbitrary constants. We refer  to this as the singleton action on the basis that if treated by itself, the field $B_2$ can be eliminated to yield the well known $N=1$ supersymmetric interacting supersingleton action. 

In summary, we have the full renormalized action
\begin{center}
\be
\boxed{
\begin{aligned}
S_{\rm ren} &= S_{\rm reg} + S_{{\rm ct},1} + S_{{\rm ct},2} + S_{\it singleton}
\label{b4}
\end{aligned}}
\ee
\end{center}
which is finite and invariant under the ordinary supersymmetry transformation,
\be
\delta_{\eta_-} S_{\rm ren}=0\ .
\ee
In contrast, $\eta_{+}$ special supersymmetry cannot be maintained but rather (\ref{b3}) yields
\be
\delta_{\eta_+} S_{\rm ren}=
{\cal A}_{\eta_+} \equiv
\int d^3 x\, \overline{\eta_+}\left(
A_1\,\slashed{\partial} \chi_{1+}-i B_2\,\gamma^5\,\chi_{1+} \right)
\;,
\ee
where $\lambda_D= \overline{\eta_-} \eta_+$. By Wess-Zumino consistency condition, this anomaly implies the presence of a dilatation anomaly as follows:
\be
\delta_{\eta_-} {\cal A}_{\eta_+}- \delta_{\eta_+} {\cal A}_{\eta_-}=\delta_{\lambda_D} S_{\rm ren} = {\cal A}_{\lambda_D}\ 
\ee
Since ${\cal A}_{\eta_-}= \delta_{\eta_-}S_{\rm ren}=0$, it follows that
\be
{\cal A}_{\lambda_D}= \int d^3x\,\lambda_D \,\Big( - \partial_a A_1 \partial^a A_1 +  B_2^2 -\frac12 \overline\chi_{1+} \slashed{\partial} \chi_{1+}\Big) \ .
\ee
The bosonic part of this result agrees with that of \cite{Petkou:1999fv,deHaro:2000xn}, where the anomaly is computed for scalar field in fixed $AdS$ background with action $\int d^4x \left(\partial_\mu \phi \partial^\mu \phi+m^2 \phi^2\right)$ for $m^2= k^2-\frac94$ with $k=0,1,2, \dots$. Note that $\lambda_D$ is not constant, but its derivative is proportional to the conformal boost parameter as follows $\partial_a \lambda_D= \overline\eta_{1+}\gamma_a \eta_{2+} \equiv \lambda_{K_i}$. For constant $\lambda_D$, ${\cal A}_{\lambda_D}$  is proportional to the free supersingleton action in which both $\eta_+$  and $\eta_-$ symmetries are present. This is consistent with the Wess-Zumino consistency condition
\be
\delta_{\eta_-}{\cal A}_{\lambda_D} -\delta_{\lambda_D}{\cal A}_{\eta_-}= {\cal A}_{\eta_-^{\rm comp}} =0\ .
\ee

\subsection{General variation and boundary conditions}
%%%%%%%%%%%%%%%%%%%%%%%%%%%%%%%%%%%%%%%%%%%%%%%%%%%%%%%%%%%%%%%%%%%%%%%%%%%%%%%%%%%%%

We now turn to the general variation of $S_{\rm ren}$. 
Starting from (\ref{gv}) and taking into account the contributions from
the counterterms $S_{{\rm ct},1}$, $S_{{\rm ct},2}$, we obtain
\be
\delta S_{\rm ren} = \int d^3 x \; \Big\{ -4\,{\mathfrak A}_2 \delta A_1+2\,B_1 \delta B_2 +\overline{\chi_{2-}} \delta\chi_{1+}\Big\}
\ .
\label{genvar}
\ee
Thus we need to impose either Dirichlet boundary conditions
\be
\mbox{Dirichlet:}\quad \delta A_1=0\ ,\quad \delta B_2=0\ ,\quad \delta \chi_{1+}=0\ ,
\ee
which form a supersymmetric set, or the Neumann boundary conditions
\be
\mbox{Neumann:}\quad {\mathfrak A}_2=0\ ,\quad B_1=0\ ,\quad \chi_{2-}=0\ .
\label{nbc}
\ee
which transform into each other under $\eta_- $ supersymmetry but not the $\eta_+ $ special supersymmetry. 

In the framework of AdS/CFT correspondence, imposing the Dirichlet boundary conditions, the boundary values of the bulk fields couple to dimension $5/2$ operators to be built from the boundary CFT. We are not in a position to specify such a CFT at this point since there are issues regarding the definition of the norms of the singletonic states, as we shall discuss further in section 7. As is well known \cite{Klebanov:1999tb}, there is also an alternative quantization scheme in which the dimensions of the boundary fields and the operators coupled to are interchanged, and this is achieved by performing a Legendre transformation of the renormalized action, which we discuss next.

\subsection{Legendre transformation}
%%%%%%%%%%%%%%%%%%%%%%%%%%%%%%%%%%%%%%%%%%%%%%%%%%%%%%%%%%%%%%%%%%%%%%%%%%%%%%%%%%%%%

From (\ref{genvar}), we find that the Legendre transformed action is given by
\begin{equation}
    S_{\rm L} =
    S_{\rm ren} - \int d^3x \left\{
    \frac{\partial{\cal L}}{\partial A_1}\,A_1
    +\frac{\partial{\cal L}}{\partial B_2}\,B_2
    +\frac{\partial{\cal L}}{\partial \chi_{1+}}\chi_{1+}
    \right\}
    =
    S_{\rm ren} 
+   2\,S'
    \;,
\end{equation}
with $S'$ from (\ref{Sf+}).
With the above result, we find that the Legendre transformed action is also 
invariant under ordinary supersymmetry transformation
\be
\delta_{\eta_-} S_{\rm L}=0\ ,
\ee
however also not invariant under special supersymmetry $\delta_{\eta_+}$.

Note that $S_{\rm ren}+S_{\rm L}$ is invariant under both, ordinary and special supersymmetry. It is also worth noting that the full supersymmetry of $S_L$ is spoiled by the factor of two in front of $S'$, because the combination $S_{\rm ren}+S'$ is fully supersymmetric.

Let us now write $S_{\rm L}$ explicitly. It is given by
\bea
S_{\rm L} &=& \int_\varepsilon^\infty \! dr \int dx^3\,{\cal L}
+\frac12\, \int dx^3\,\sqrt{-\gamma}\,\left(
A^2+A^4+3\,B^2+\frac{2\,B^2}{{\rm log}\,\varepsilon}
+2\,{\rm log}\,\varepsilon\,\gamma^{mn}\,\partial_m A \partial_n A
\right)
\nn\w2
&& +\int d^3 x\, \sqrt{-\gamma}\, 
\Big\{ 
\frac13\, A^6
+\frac14\,\big(
\overline{\chi}\chi+2\,({\rm log} \,\varepsilon)\,
\overline{\chi}\gamma^m\partial_m \chi
\big)
\Big\}
\nn\w2
&& 
+\int d^3 x\, \Big( -\overline{\chi_{2-}}\chi_{1+}
+4A_1{\mathfrak A}_2-2B_1 B_2\Big)\ .
\eea
The general variation of this action gives
\be
\delta S_{\rm L} = \int d^3 x \; \Big\{ -4\,(\delta {\mathfrak A}_2)  A_1+2\,(\delta B_1) B_2 +(\delta \overline{\chi_{2-}} )\chi_{1+}\Big\}
\ .
\label{genvar2}
\ee
Choosing the boundary conditions \eq{nbc} now implies that the boundary values of the fields will couple to dimension $1/2$ operators on the boundary CFT. However, as mentioned above, while  these boundary conditions are invariant under ordinary supersymmetry, they break the special supersymmetry. The nature of a possible boundary CFT in this setting remains an open problem, 

\section{Comments}
%%%%%%%%%%%%%%%%%%%%%%%%%%%%%%%%%%%%%%%%%%%%%%%%%%%%%%%%%%%%%%%%%%%

Taking the results above as a starting point for computation of correlation functions entail a number of obstacles. To begin with, the finiteness and conservation of asymptotic symmetries need to be established. This entails a proper definition of norms with respect to which all fields have finite norms. Given that the holographically renormalized actions involves boundary terms engineered to remove unwanted singularities in the action near the boundary, the definition of the norm becomes a subtle problem for which we are not aware of a universal solution. 
If one merely studies the solution of the free field equations, in the case of the $B$-field for which $E_0=3/2$,  one finds a solution which has finite norm defined by $(\phi_1,\phi_2) =i\int d^3x\,\sqrt{-g} g^{0\n} (\phi_1\overleftrightarrow{\partial_\n} \phi_2)$.  For a field with $E_0=1/2$ , however, this norm diverges, as shown in \cite{Starinets:1998dt}, where a definition of the norm modified by a multiplication with the factor $E_0-\frac12$ is shown to give a finite answer. 

These discussions of norms do not take into account the boundary effects encoded in the holographically renormalized action. The consequences of coupling scalar fields with $E_0=d/2$ to $AdS_{d+1}$ gravity were studied in \cite{Henneaux:2004zi}, where the existence of finite and conserved asymptotic charges was analyzed. It was found that while the scalar field contributes logarithmic divergent piece, it is cancelled by gravitational contribution which results from the fact that the logarithmic fall of the scalar field implies also logarithmically falling part to the metric, whose contribution to the asymptotic charge formula conspires with that of the scalar field. The implications of this result for a proper definition of a norm seems to require further investigation.

In the case of scalar field with $E_0=1/2$, the definition of norms and its consequences has been considered in  \cite{Andrade:2011dg,Ohl:2012bk}. In \cite{Andrade:2011dg}, a renormalized norm is defined which is finite for time-like 3-momentum, i.e. ${\vec p}^2<0$ , but diverges for light-light 3-momentum, i.e. ${\vec p}^2=0$. Furthermore, the norm is negative definite for ${\vec p}^2>0$, which implies a ghostly tachyonic state. 

In an alternative approach, in \cite{Ohl:2012bk}, a finite boundary action proportional to $\int \kappa (\partial_i A)^2$ is added to renormalized action. Next, in a renormalized definition of the norm, the field $A$ is redefined as $A'= \kappa^{-1/2} A$, and the limit $\kappa \to \infty$ is taken. Even though  different values of $\kappa$ are related by an AdS transformation and therefore should be equivalent, this is not necessarily so since dilatation invariance is broken. Taking this limit is argued to imply that $A$ is a free singleton field on the boundary, and no interactions can arise. Whether an admissible similar mechanism can be formulated in the context of the interacting singleton action we have considered (see \eq{SL}) and its full consequences remain to be investigated.

There exists an alternative formulation of singleton field theory in which the field equation $(\Box-\frac54 )A=0$ is replaced by $(\Box-\frac54 )^2A=0$ \cite{Flato:1986uh}. It was shown in \cite{Flato:1986uh} that a ghost state resulting in this way provides the gauge mode in the sense described in \cite{Flato:1980we}, and it is argued that this approach furnishes a better alternative to the quantization of the singleton. In the AdS/CFT correspondence context, it was later shown that this bulk description of the singleton gives rise to a logarithmic conformal field theory on the boundary of $AdS_4$ \cite{Kogan:1999bn}. 

 In this paper we have focused on the holographic quantization of the global limit of the singleton coupled to bulk supergravity. This limit already captures the crucial issues involved. Once the highlighted obstacles are overcome, it would be a relatively straightforward matter to include the supergravity fields in this analysis along the lines for ordinary scalars coupled to $N=1$ supergravity in \cite{Amsel:2009rr}. Following \cite{Freedman:2016yue}, we expect that the boundary terms obtained in the global limit are not changed by reanalysis at the level of $N = 1$ supergravity. 

The analysis presented in this paper may be carried out for the $N>1$ supersingletons in $4D$. In particular the case of $N=8$ supersingleton is of considerable interest in the context of 7-sphere compactification of $11D$ supergravity \cite{Sezgin:1983ik,Casher:1984ym,Nilsson:2018lof}. It would also be interesting to investigate a bulk description of the singleton-like representations of the $AdS$ supergroup in dimensions $D\le 7$. In the case of half-maximal $AdS_7$ supergroup, the singleton representation is an $N=(1,0), 6D$ tensor multiplet admitting a conformal field theoretic description on the boundary \cite{Nicolai:1988ek}. Considering only the 2-form potential in this multiplet, its $7D$ bulk description has been described as a particular BF theory in \cite{Maldacena:2001ss}. It would be interesting to generalize this construction for the case of the full supersingleton multiplet, and investigate its coupling  to the half-maximal gauged supergravity in $7D$ coupled to vector multiplets.

\section*{Acknowledgement}
%%%%%%%%%%%%%%%%%%%%%%%%%%%%%%

We are grateful to Don Marolf and Christoph Uhlemann for helpful comments.  We thank each other's home institutions for hospitality during this work. The work of E.S. is supported in part by NSF grants PHYS-2112859 and PHYS-2413006. 

%\bibliographystyle{utphys}
%\bibliography{refs}

\begin{thebibliography}{10}

\bibitem{Nicolai:1988ek}
H.~Nicolai, E.~Sezgin, and Y.~Tanii, ``Conformally invariant supersymmetric
  field theories on {$S^p\times S^1$} and super $p$-branes,''
  \href{http://dx.doi.org/10.1016/0550-3213(88)90077-6}{{\em Nucl. Phys. B}
  {\bfseries 305} (1988) 483--496}.

\bibitem{Flato:1980we}
M.~Flato and C.~Fronsdal, ``Quantum field theory of singletons: {T}he {R}ac,''
  \href{http://dx.doi.org/10.1063/1.524993}{{\em J. Math. Phys.} {\bfseries 22}
  (1981) 1100}.

\bibitem{Starinets:1998dt}
A.~Starinets, ``Singleton field theory and {F}lato-{F}ronsdal dipole
  equation,'' \href{http://dx.doi.org/10.1023/A:1007644223085}{{\em Lett. Math.
  Phys.} {\bfseries 50} (1999) 283--300},
  \href{http://arxiv.org/abs/math-ph/9809014}{{\ttfamily
  arXiv:math-ph/9809014}}.

\bibitem{Breitenlohner:1982jf}
P.~Breitenlohner and D.~Z. Freedman, ``Stability in gauged extended
  supergravity,'' \href{http://dx.doi.org/10.1016/0003-4916(82)90116-6}{{\em
  Annals Phys.} {\bfseries 144} (1982) 249}.

\bibitem{Sezgin:2020avr}
E.~Sezgin, ``{11D supergravity on $AdS_4 \times S^7$ versus $AdS_7 \times
  S^4$},'' \href{http://dx.doi.org/10.1088/1751-8121/ab8e67}{{\em J. Phys. A}
  {\bfseries 53} no.~36, (2020) 364003},
  \href{http://arxiv.org/abs/2003.01135}{{\ttfamily arXiv:2003.01135
  [hep-th]}}.

\bibitem{Fischbacher:2009cj}
T.~Fischbacher, ``Fourteen new stationary points in the scalar potential of
  ${SO}(8)$-gauged ${{\cal N}}=8, {D}=4$ supergravity,''
  \href{http://dx.doi.org/10.1007/JHEP09(2010)068}{{\em JHEP} {\bfseries 09}
  (2010) 068}, \href{http://arxiv.org/abs/0912.1636}{{\ttfamily arXiv:0912.1636
  [hep-th]}}.

\bibitem{Fischbacher:2010ec}
T.~Fischbacher, K.~Pilch, and N.~P. Warner, ``New supersymmetric and stable,
  non-supersymmetric phases in supergravity and holographic field theory,''
\href{http://arxiv.org/abs/1010.4910}{{\ttfamily arXiv:1010.4910 [hep-th]}}.
%%CITATION = ARXIV:1010.4910;%%.

\bibitem{Cremmer:1982en}
E.~Cremmer, S.~Ferrara, L.~Girardello, and A.~Van~Proeyen, ``Yang-{M}ills
  theories with local supersymmetry: {L}agrangian, transformation laws and
  super{H}iggs effect,''
\href{http://dx.doi.org/10.1016/0550-3213(83)90679-X}{{\em Nucl. Phys.}
  {\bfseries B212} (1983) 413}.
%%CITATION = NUPHA,B212,413;%%.

\bibitem{Freedman:2012zz}
D.~Z. Freedman and A.~Van~Proeyen, {\em {Supergravity}}.
\newblock Cambridge University Press,
2012.
\newblock
%%CITATION = INSPIRE-1123253;%%.

\bibitem{Amsel:2009rr}
A.~J. Amsel and G.~Comp\`ere, ``Supergravity at the boundary of {AdS}
  supergravity,'' \href{http://dx.doi.org/10.1103/PhysRevD.79.085006}{{\em
  Phys. Rev. D} {\bfseries 79} (2009) 085006},
  \href{http://arxiv.org/abs/0901.3609}{{\ttfamily arXiv:0901.3609 [hep-th]}}.

\bibitem{Freedman:2016yue}
D.~Z. Freedman, K.~Pilch, S.~S. Pufu, and N.~P. Warner, ``Boundary terms and
  three-point functions: {A}n {AdS/CFT} puzzle resolved,''
  \href{http://dx.doi.org/10.1007/JHEP06(2017)053}{{\em JHEP} {\bfseries 06}
  (2017) 053}, \href{http://arxiv.org/abs/1611.01888}{{\ttfamily
  arXiv:1611.01888 [hep-th]}}.

\bibitem{Bianchi:2001kw}
M.~Bianchi, D.~Z. Freedman, and K.~Skenderis, ``Holographic renormalization,''
  \href{http://dx.doi.org/10.1016/S0550-3213(02)00179-7}{{\em Nucl. Phys. B}
  {\bfseries 631} (2002) 159--194},
  \href{http://arxiv.org/abs/hep-th/0112119}{{\ttfamily arXiv:hep-th/0112119}}.

\bibitem{Petkou:1999fv}
A.~Petkou and K.~Skenderis, ``{A Nonrenormalization theorem for conformal
  anomalies},'' \href{http://dx.doi.org/10.1016/S0550-3213(99)00514-3}{{\em
  Nucl. Phys. B} {\bfseries 561} (1999) 100--116},
  \href{http://arxiv.org/abs/hep-th/9906030}{{\ttfamily arXiv:hep-th/9906030}}.

\bibitem{deHaro:2000xn}
S.~de~Haro, S.~N. Solodukhin, and K.~Skenderis, ``{Holographic reconstruction
  of spacetime and renormalization in the AdS/CFT correspondence},''
  \href{http://dx.doi.org/10.1007/s002200100381}{{\em Commun. Math. Phys.}
  {\bfseries 217} (2001) 595--622},
\href{http://arxiv.org/abs/hep-th/0002230}{{\ttfamily arXiv:hep-th/0002230}}.
%%CITATION = HEP-TH/0002230;%%.

\bibitem{Klebanov:1999tb}
I.~R. Klebanov and E.~Witten, ``{AdS / CFT correspondence and symmetry
  breaking},'' \href{http://dx.doi.org/10.1016/S0550-3213(99)00387-9}{{\em
  Nucl.Phys.} {\bfseries B556} (1999) 89--114},
\href{http://arxiv.org/abs/hep-th/9905104}{{\ttfamily arXiv:hep-th/9905104
  [hep-th]}}.
%%CITATION = HEP-TH/9905104;%%.

\bibitem{Henneaux:2004zi}
M.~Henneaux, C.~Martinez, R.~Troncoso, and J.~Zanelli, ``Asymptotically anti-de
  {S}itter spacetimes and scalar fields with a logarithmic branch,''
  \href{http://dx.doi.org/10.1103/PhysRevD.70.044034}{{\em Phys. Rev. D}
  {\bfseries 70} (2004) 044034},
  \href{http://arxiv.org/abs/hep-th/0404236}{{\ttfamily arXiv:hep-th/0404236}}.

\bibitem{Andrade:2011dg}
T.~Andrade and D.~Marolf, ``{AdS/CFT} beyond the unitarity bound,''
  \href{http://dx.doi.org/10.1007/JHEP01(2012)049}{{\em JHEP} {\bfseries 01}
  (2012) 049}, \href{http://arxiv.org/abs/1105.6337}{{\ttfamily arXiv:1105.6337
  [hep-th]}}.

\bibitem{Ohl:2012bk}
T.~Ohl and C.~F. Uhlemann, ``Saturating the unitarity bound in {AdS/CFT$_{\rm
  (AdS)}$},'' \href{http://dx.doi.org/10.1007/JHEP05(2012)161}{{\em JHEP}
  {\bfseries 05} (2012) 161}, \href{http://arxiv.org/abs/1204.2054}{{\ttfamily
  arXiv:1204.2054 [hep-th]}}.

\bibitem{Flato:1986uh}
M.~Flato and C.~Fronsdal, ``The singleton dipole,''
  \href{http://dx.doi.org/10.1007/BF01212320}{{\em Commun. Math. Phys.}
  {\bfseries 108} (1987) 469}.

\bibitem{Kogan:1999bn}
I.~I. Kogan, ``Singletons and logarithmic {CFT} in {AdS / CFT}
  correspondence,'' \href{http://dx.doi.org/10.1016/S0370-2693(99)00576-6}{{\em
  Phys. Lett. B} {\bfseries 458} (1999) 66--72},
  \href{http://arxiv.org/abs/hep-th/9903162}{{\ttfamily arXiv:hep-th/9903162}}.

\bibitem{Sezgin:1983ik}
E.~Sezgin, ``The spectrum of the eleven-dimensional supergravity compactified
  on the round seven sphere,''
  \href{http://dx.doi.org/10.1016/0370-2693(84)91872-0}{{\em Phys. Lett. B}
  {\bfseries 138} (1984) 57--62}.

\bibitem{Casher:1984ym}
A.~Casher, F.~Englert, H.~Nicolai, and M.~Rooman, ``The mass spectrum of
  supergravity on the round seven sphere,''
\href{http://dx.doi.org/10.1016/0550-3213(84)90392-4}{{\em Nucl. Phys.}
  {\bfseries B243} (1984) 173}.
%%CITATION = NUPHA,B243,173;%%.

\bibitem{Nilsson:2018lof}
B.~E.~W. Nilsson, A.~Padellaro, and C.~N. Pope, ``The role of singletons in
  {S$^{7}$} compactifications,''
  \href{http://dx.doi.org/10.1007/JHEP07(2019)124}{{\em JHEP} {\bfseries 07}
  (2019) 124}, \href{http://arxiv.org/abs/1811.06228}{{\ttfamily
  arXiv:1811.06228 [hep-th]}}.

\bibitem{Maldacena:2001ss}
J.~M. Maldacena, G.~W. Moore, and N.~Seiberg, ``D-brane charges in five-brane
  backgrounds,'' \href{http://dx.doi.org/10.1088/1126-6708/2001/10/005}{{\em
  JHEP} {\bfseries 10} (2001) 005},
  \href{http://arxiv.org/abs/hep-th/0108152}{{\ttfamily arXiv:hep-th/0108152}}.

\end{thebibliography}

\providecommand{\href}[2]{#2}\begingroup\raggedright\endgroup

\end{document}